# Comment on "Spin-rotation coupling in non-exponential decay of hydrogenlike heavy ions" by G. Lambiase et al.

Thomas Faestermann[a)]

*Physik Department E12, Technische Universität München, D85748 Garching, Germany*

I note that the explanation of a recently posted paper by Lambiase et al. [1] for the oscillations in the electron capture decay rate, reported by Litvinov et al.[2], requires the cancellation of a difference by eleven orders of magnitude for the two independent measurements. Such a cancellation is (impossible)².

Recently measurements at the storage ring ESR of GSI were reported [2] on electron capture (EC) decay of hydrogen-like $^{140}Pr^{58+}$ and $^{142}Pm^{60+}$ ions, where superimposed on the exponential decay an oscillatory behaviour with a frequency of about 0.14 $Hz$ was observed. Lambiase et al. [1] suggest as possible explanation the Thomas precession of the electron spin relative to the nuclear spin such that periodically the hyperfine state with total spin F=3/2 is populated, for which an EC decay to the daughter state with F=1/2 is forbidden. They derive a frequency $\Omega/2\pi$ for this precession as

$$\Omega = -\left[a_e - 1 + \frac{2}{\gamma_e} + \left(\tilde{\mu} - \frac{2Z}{A}\right)\frac{m_e}{m_p}\right] \cdot \frac{eB}{m_e} \quad (1).$$

Here $a_e = (|g_e| - 2)/2$ is the electron magnetic moment anomaly, $\gamma_e$ is the Lorentz factor of the bound electron, $\tilde{\mu}$ is the magnetic moment of the nucleus in units of the nuclear magneton, Z and A are atomic and mass number of the nucleus, $m_e$ and $m_p$ are the electron and proton mass resp., and B is the magnetic field of the ESR averaged over the circumference. B is obtained as $B\rho \cdot 2\pi/L_{ESR}$, $L_{ESR}$ is the circumference and correctly quoted as 108.3$m$, but the magnetic rigidity of the bending magnets $B\rho$ is a little different from what is quoted in [1]. It can be estimated with their numbers:

$B\rho = \beta\gamma \cdot Mc/Q \approx 0.71 \cdot 1.43 \cdot 140 \cdot 931.5 \, MeV/c/58e = 7.61 \, Tm$

for $^{140}$Pr e.g. (the actually measured value was 7.69$Tm$), with the velocity $\beta=v/c$ and the Lorentz factor $\gamma$ of the circulating ion that has mass M and charge Q. Thus $B \approx 0.441 T$ and the second factor of $\Omega$ in equation (1) is $eB/m_e \approx 7.77 \cdot 10^{10} \, rad/s$.

Now, to arrive at an oscillation frequency $\Omega/2\pi \approx 0.14 Hz$, as reported by [2], Lambiase et al. are courageous enough to tune the first factor in equation (1) such that the difference in the bracket is cancelled to about $10^{-11}$. They not only require such a cancellation for one case but for both cases, $^{140}$Pr and $^{142}$Pm, by choosing the relativistic factor $\gamma_e$ of the bound electron accordingly. It is remarkable that instead of quoting eleven significant digits, they quote only six. At least they remark that the nuclear magnetic moments for the two nuclei should be better known than the estimate of [2] with two significant digits. Moreover, the values of $\gamma_e \approx 1.88$ dropping out of equation (1) correspond to a kinetic energy of $(\gamma_e - 1)m_e c^2 \approx 450 \, keV$ and a potential energy of –690 $keV$ at the radius of 123 $fm$. How that should be compatible with the total binding energy of –49.6 $keV$ for $^{140}Pr^{58+}$, calculated from the Dirac equation [3], they do not care.

As a conclusion, I am grateful that peer reviewed journals still exist.
Thanks to Reiner Krücken for critically reading this manuscript.

---

[a)] thomas.faestermann@ph.tum.de